# One-dimensional ghost imaging with an electron microscope: a route towards ghost imaging with inelastically scattered electrons

E. Rotunno[1], S. Gargiulo[2], G. M. Vanacore[3], C. Mechel[4], A. H. Tavabi[5], R. E. Dunin-Borkowski[5], F. Carbone[2], I Maidan[2], M. Zanfrognini[6], S. Frabboni[6], T. Guner[7], E. Karimi[7], I. Kaminer[4] and V. Grillo[1*]

1 Centro S3, Istituto di Nanoscienze-CNR, 41125 Modena, Italy

2 Institute of Physics, Laboratory for Ultrafast Microscopy and Electron Scattering (LUMES), École Polytechnique Fédérale de Lausanne, Station 6, Lausanne, 1015, Switzerland

3 Department of Materials Science, University of Milano-Bicocca, Via Cozzi 55, 20121, Milano, Italy

4 Department of Electrical Engineering, Technion-Israel Institute of Technology, Haifa 32000, Israel

5 Ernst Ruska-Centre for Microscopy and Spectroscopy with Electrons and Peter Grünberg Institute, Forschungszentrum Jülich, 52425 Jülich, Germany

6 Dipartimento FIM, Universitá di Modena e Reggio Emilia, 41125 Modena, Italy

7 Department of Physics, University of Ottawa, 25 Templeton Street, Ottawa, Ontario K1N 6N5, Canada

*email: vincenzo.grillo@nano.cnr.it

**In quantum mechanics, entanglement and correlations are not sporadic curiosities but ubiquitous phenomena that are at the basis of interacting quantum systems. In electron microscopy, such concepts have not yet been explored extensively. For example, inelastic scattering can be reanalyzed in terms of correlations between an electron beam and a sample. Whereas classical inelastic scattering results in a loss of coherence in the electron beam, a joint measurement performed on both the electron beam and the sample excitation could restore the coherence and "lost information." Here, we propose to exploit joint measurement in electron microscopy to achieve a counterintuitive application of the concept of ghost imaging, which was first proposed in quantum photonics. The same approach can be applied partially in electron microscopy by performing a joint measurement between a portion of the transmitted electron beam and a photon emitted from the sample, which arrives at a bucket detector, thereby allowing the formation of a one-dimensional virtual image of an object that has not interacted directly with the electron beam. The technique is of particular interest for low-dose imaging of electron-beam-sensitive materials using a minimal radiation dose, as the object interacts with other forms of waves, such as photons or surface plasmon polaritons, rather than with the electron beam itself. We provide a theoretical description of the concept for inelastic electron-sample interactions, in which an electron excites a single quantum of a collective mode, such as a photon, a plasmon, a phonon, a magnon or any optical polariton. The excited collective mode remains correlated with the interacting electron. In optical ghost imaging, spatially-entangled photon pairs or classically-correlated photons are used to extract information (such as phase, transmission**

**and birefringence) from an object through a joint measurement. In electron microscopy, a similar approach can be applied to highly correlated systems, such as electron-surface plasmon polaritons, to obtain additional information about the sample. In addition to applications in materials science, the approach can be used to shape electron wavefunctions by post-selecting a particular correlated collective mode. As long as the generated collective mode states are not determined, the quantum state of the electron beam is a maximally mixed state. The approach could be used to broaden the range of applications of electron microscopy, as it enables a sample to be imaged with the "eyes" of a different probe, e.g., a plasmon polariton that originates in the material. In a broader sense, aside from achieving ghost or interaction-free imaging, the possibility to exploit quantum coherent effects is important for developing quantum imaging techniques in electron microscopy.**

One the most elusive phenomena in quantum mechanics is commonly referred to as the "collapse of the wavefunction." This nomenclature describes the fact that a non-unitary transformation of the wavefunction during the measurement results in a loss of quantum information while it interacts with the measuring system. For instance, when a quantum particle interacts with a macroscopic measuring device, the quantum particle and the measuring device become fully correlated in their interacting degrees of freedom, while other quantum features of the system (i.e., the particle and the measuring device) are unaffected. Therefore, tracing over the measuring device's degrees of freedom results in a loss of information about the device and particle quantum state, which is referred to as a loss of coherence[1,2]. This loss of coherence and information is an intrinsic aspect of the measurement problem, as well as the collapse of the wavefunction. Recovery of the coherence and quantum information requires a tomographic measurement on a desired number of quantum system ensembles[3],[4].

In electron microscopy, the inelastic interaction of electrons with matter provides an ideal model system for testing the phenomenology of loss of coherence. This problem has already been investigated in many studies[5,6,7,8,9,10,11,12,13]. Inelastic scattering is usually observed using electron energy-loss measurements and offers a variety of possibilities in terms of position localization of the wavefunction[14]. Of particulat interest for our work are interactions in the low energy-loss region, where the excitations include collective modes, such as plasmons, polaritons, magnons and phonons, which have relatively long lifetimes for propagation in the sample. A (joint) measurement performed on the collective modes and the electrons provides a unique way to infer the state of the sample, which is usually considered to be a classical environment for the electron probe. The present work utilizes such joint measurements, instead of simply considering the loss of coherence of the electron wavefunction following inelastic scattering. By considering the sample and the probe as a correlated system[15], one can extract more information about the sample and the probe by performing a joint measurement. We obtain the counter-intuitive result that, under appropriate conditions, an image of part of the sample is formed on an electron detector although the electron never interacted with that portion of the sample, which has been explored by travelling collective modes that are correlated with the electron.

Ghost imaging is an idea based on the concept of joint measurement[16], which originally emerged from studies of the Einstein-Podolsky-Rosen (EPR) paradox[17,18,19,20]. Subsequently, it was proved that classically-correlated states, through raster scanning and joint measurements, can also provide ghost imaging[21]. As the superposition principle allows a quantum system to be in different eigenstates simultaneously, many outcomes of the evolution of a given quantum state are possible. However, when the outcome of a classical measurement is defined, one specific evolution is singled out and it becomes the "real" one as a result of decoherence or simply state post-selection. When two quantum particles (in our case an electron and a collective mode) are entangled, their quantum state is not factorized. Therefore, a measurement performed on of the particles can reveal the other particle's state, giving rise to the seemingly paradoxical effects of backward causality[22,23]. In practice, the entanglement enables a measurement of the sample through the joint measurement.

The entanglement translates into a well-defined optical scheme, referred to as quantum "ghost imaging", which we apply here to the electron case. In optical "ghost imaging"[24,25], an entangled photon pair (or classically correlated photons) is sent along two different paths. One follows a path where both the object and a bucket detector are located consecutively, while the other is sent to a spatially-resolved detector such as CCD camera. The CCD camera is activated only when the bucket detector is hit by the twin photon in coincidence, resulting in the formation of an image based on this joint measurement on the CCD camera. The counter-intuitive result is that an image of the sample is formed on the CCD camera even if such a photon never passed through the sample. This phenomenon can be explained by considering time inversion of the rays emanating from the bucket, going through the sample/object and reaching the bifurcation where the information reaches the second photon.

In order to carry out a similar experiment with electrons, one can utilize the excitation of surface plasmon polaritons (SPPs) in electron scattering and consider how a joint measurement between the SPP and electron states can give rise to an image on the electron beam from the SPP evolution. The electron is detected using a spatially-resolved camera similar to its optical counterpart, while the SPPs traverse through the sample, are converted to photons by means of a localized grating and are detected by a bucket detector. The signal from the bucket detector (SPPs) triggers the camera, resulting in the formation of a ghost image on the electron beam's side.

The coherent dynamics of collective modes in electron microscopy have previously been studied using ultrafast TEM (UTEM)[26,27]. This technique involves the use of femtosecond laser excitation of the sample synchronized with a pulsed electron probe. The laser-driven electron sources that are needed for such experiments are limited in their coherence and brightness. Here, we propose a complementary approach that relies on a joint measurement scheme based on correlation between the collective modes and the electron, which can be achieved with cathodoluminescence. Our approach can be regarded as a time reversal of a pump-probe process.

We consider a thin slab of metal that can support the propagation of SPPs, with the electron beam propagating along the *z* direction, which is perpendicular to the slab. The SPPs lie in the *x-y* plane. In the *x* direction, a discontinuous obstacle that takes the form of a double slit is inserted[28]. A more general version of this *gedanken* experiment can be obtained by exchanging the slits by objects that affect the SPP amplitude and phase.

d

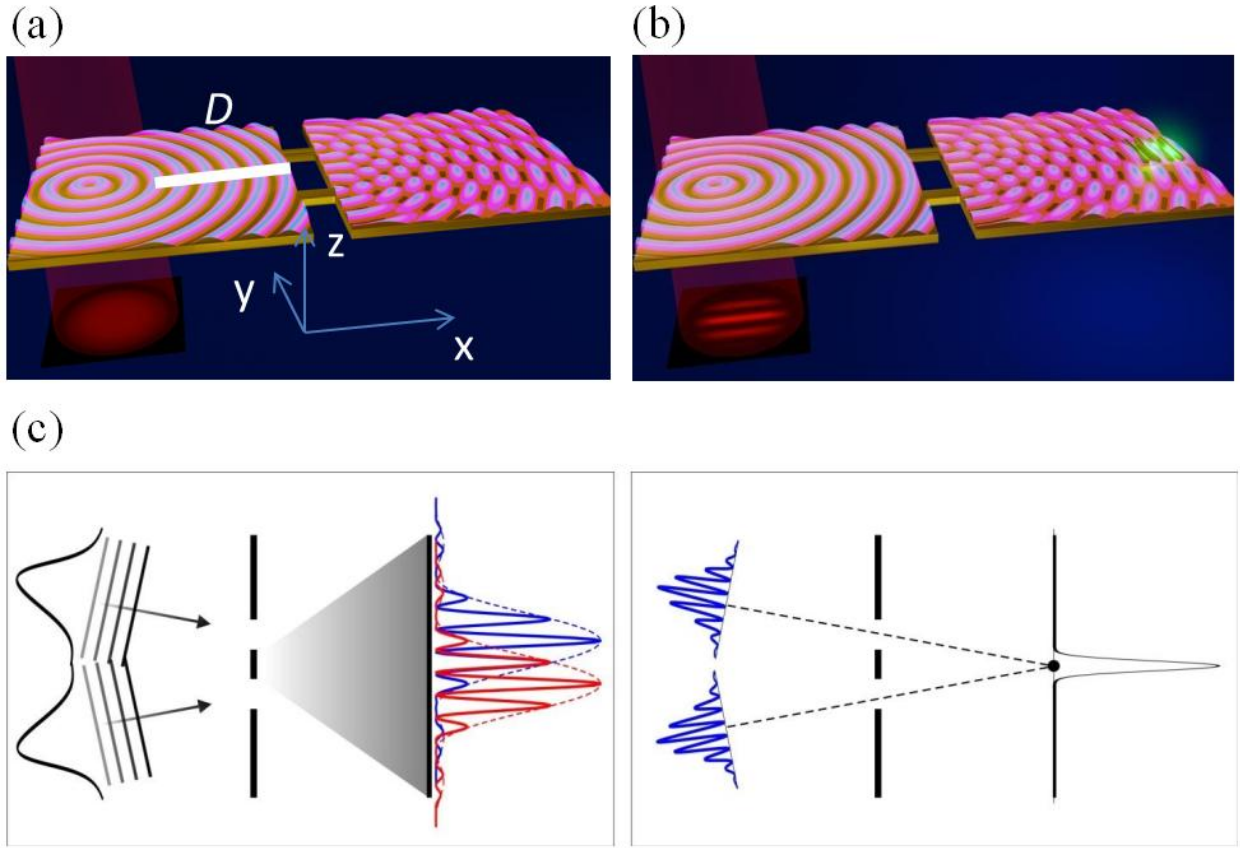

*Figure 1.* Schematic diagram illustrating the "ghost imaging" experiment. (a) The electron beam impinges on a metallic slab and produces an SPP propagating along its surface. The SPP is scattered by a double slit feature and produces an incoherent sum of many interference patterns, one for each SPP momentum. The image formed on the CCD camera is a truncated Gaussian. (b) Conversion of the SPP to a photon using a suitable optical grating defines the joint wavefunction between the specific SPP mode and the electron state imaged in the CCD detector. The image formed by the joint measurement on the electron beam side is now a diffraction pattern from the double slit. (c) The left panel shows the propagation and interference of two forward wave components with two particular wavevectors extracted from all of the wavelet components associated with the generated SPP. The right panel shows the reverse wave components of the SPP associated with the point source, representing the corresponding coincident photon scattered from the grating.

In this experiment, the SPP is generated by electrons in the left part of the slab (Fig. 1a) and is converted into light through a grating that is positioned in the right part of the slab (Fig. 1b). The grating enables the SPP to be outcoupled by breaking the translational symmetry and adding a momentum proportional to the grating spatial frequency - such that each quantum of SPP is transformed into a free-space propagating photon[29]. The grating is responsible for the spatial localization and energy selectivity of the joint measurement. The grating is confined only in a specific region of lateral size Δy, which allows the sample to be resolved in the y direction with a resolution that is based on the features encoded on the correlation between the propagated SPP and the scattered electrons. Each spatially-resolved point in the *y* direction is expected to contain integrated information along the *x* direction, such as thickness (i.e., *x* extension). If a technique such as cathodoluminescence is used to detect the emitted photon, then the grating can be considered as a bucket detector and the two slits as the sample. We infer the image of the two slits or the image of the Fraunhofer interference pattern, which should be formed on the CCD for different defoci that correspond to conjugate planes of the electron propagation after the interaction.

In order to demonstrate the feasibility of the approach and to infer its limits, we start from a description of the SPP in quantum mechanical terms. In a ghost imaging scheme, we need to specify the correlation in a specific basis. Here, we assume that the overall in-plane momentum of the electrons and the SPP must be conserved. The joint wavefunction of the SPP and the scattered electron can then be written

$$|\Psi\rangle = \sum_{\bar{k},w} a_{\bar{k}} \, E^{\text{FWD}}(\bar{k},\omega) \otimes |\psi_e(-\bar{k}, \varepsilon - \hbar\omega))\rangle \,, \qquad (1)$$

where $E^{\text{FWD}}(\bar{k})$ is a specific component of the SPP electric field and $\psi_e$ is the electron wavefunction. In Eq. 1, $\bar{k}$ is the SPP momentum and the scattered electron should recoil by exactly $-\bar{k}$. The decomposition factor $a_{\bar{k}}$ depends on the illumination conditions and on the coupling. We also wrote the energy conservation explicitly, such that the energy of the created SPP (given by $\hbar\omega$) corresponds to the energy loss experienced by the swift electron (initially at $\varepsilon$). The superscript FWD stands for "forward" and indicates that the SPP wave propagates in the metal starting from the electron injection region.

The system wavefunction $|\Psi\rangle$ represents a pure state as long as the impinging electron is a pure state. When the electron is described by a density matrix, the joint wavefunction can be generalized accordingly. In the following examples, we assume that the electron beam is monochromated on a scale of 100 meV, which is within current experimental capabilities. (In general, we only require the uncertainty in electron energy to be smaller than the energy of the collective mode). A thorough discussion about the extent to which $|\Psi\rangle$ remains a pure state after scattering is currently at the center of debate in the time-resolved community. Recent contributions show that this can be approximately the case [30,31,32,33,34]. We assume below that we can indeed create such a condition.

In Eq. 1, the SPP wave is treated using classical fields, instead of a more complete second quantization approach[23]. This does not limit the generality of our conclusions, since the SPP probability distribution is unaltered in our proposed experiment. More explicitly, we consider the excitation of a single SPP. Although there is a possibility of a coherent multiple SPP excitation[23], the experimental conditions can be tuned to reduce coupling and avoid this situation. A generalization using a "dressed" or renormalized SPP experiment would be equivalent to this formalism, as long as the generation and losses of one of the quanta other than the first are not important [35].

To a first approximation, it is sufficient to consider a comparison with the well-known photon-induced near-field electron microscopy (PINEM) process [36,37,22], whereby a laser pulse and a swift electron impinge on a sample simultaneously. In this case, the SPPs that are generated by the swift electron are neglected, as their population is much lower than those generated by the laser[38]. It is reasonable to assume that the main problem of our ghost imaging technique, when compared to direct PINEM, is the relatively small number of events. However, in comparison to PINEM, the greater brightness of a continuous cathode with respect to a pulsed photocathode, which generates significantly less electrons, provides partial compensation. We also treat the generated SPP for a well-defined energy. This can be ensured by appropriate detection of the photon energy.

Another consequence of the semiclassical description of the field is that quantization of the collective excitation is hidden. This creates a problem, since a joint measurement between the photon and the electron is directly possible only in a quantized field framework. We therefore use an *ad hoc* manipulation to highlight the contribution of each partial wave to the SPP, so that the coincident measurement arises naturally.

We start from a text-book treatment for the excitation of plasmons by swift electrons. Each electron produces an electric field [39], which can be described by the expression

$$\bar{E}_S = \frac{2e\omega}{v^2\gamma\varepsilon}\left[\frac{i}{\gamma}K_0\left(\frac{\omega\rho}{v\gamma}\right)\hat{z} - K_1\left(\frac{\omega\rho}{v\gamma}\right)\hat{\rho}\right], \qquad (2)$$

where $K_{0,1}$ are modified Bessel functions of the second kind of order 0 and 1, $v$ is the electron velocity, $\gamma$ the relativistic factor for electrons of energy $\varepsilon$, $\omega$ is the selected photon frequency, and $\hat{\rho}$ and $\hat{z}$ are unit vectors along the $\rho$ and $z$ directions of the cylindrical coordinates $(\rho, \phi, z)$. The spectrum of temporal frequencies is itself broadly distributed, but we single out a single frequency corresponding to the excitation of a specific SPP. The form of the electric field (and the related magnetic field) reflects the distribution of the generated SPP, so that the highly-peaked (diverging) function described by Eq. 2 can be considered as a virtual plasmon source.

From Eq. 2, we retrieve the characteristic size of the source distribution $s_0 = \frac{v\gamma}{\omega}$. For generality, we describe the SPP source as a more generic peak function $S_s(\rho)$ of size $s \gg s_0$, where the extended beam waist size is also taken into account. To an extent, an extended beam can still be considered as a coherent excitation source [22], and we show below that for the experiment to work relatively broad electron illumination is preferred.

According to Eq. 1, we consider $S_{kS}(\bar{\rho})$ to be a specific function of vector position $\bar{\rho}$=(x,y), which is different for each value of the momentum vector $\bar{k}$, as we assume that the entanglement/correlation is in the momentum degree of freedom. For a fixed $\bar{k}$, we take the *ansatz* for the initial wave, i.e.,

$$E(\bar{\rho}) = S_{kS}(\bar{\rho})\exp(i\bar{k}\cdot\bar{\rho}). \qquad (3)$$

We also assume that $S_{kS}(\bar{\rho})$ has a Gaussian-like distribution with a different center for the allowed values of $\bar{k}$. We set up the electron beam waist to be sufficiently broad to be larger than SPP wavelength $\lambda_{SPP}$[14]. According to Eq. 3, the source is a series of truncated "1D plane waves" (confined in the slab) originating from the injection area. The SPP is locked in a transverse magnetic (TM) mode and is treated here as a scalar wave, neglecting polarization-dependent effects.

Figure 1c (left) shows a forward-propagating SPP field $E^{\text{FWD}}(\bar{k}, \bar{\rho})$ for a given frequency, propagating from the electron injection region. Each value of $\bar{k}$ results in a different diffraction figure on the right. For the SPP, the double slit has the transfer function $T(y) = Rect_b\left(y - \frac{d}{2}\right) + Rect_b\left(y + \frac{d}{2}\right)$, where $d$ is the distance between the slits, $b$ is the size of the slits and $Rect_b$ is a support function of size $b$. The intensity of the joint measurement can be evaluated as a projection of the forward-propagating SSP $E^{\text{FWD}}(\bar{k}, \bar{\rho})$ onto the states that correspond to a certain detection outcome. The detection probability is defined by the projection of the field (in this case the SPP) on a well-defined position corresponding to the grating. Since we are solving the problem in the frequency domain, it becomes similar to a time-independent Schrödinger equation, with the solutions defined in the full domain. States with well-defined arrival points P are eigenstates of the SPP wave equation in the metal slab, since they would exist as solutions of the time-reversal Maxwell equations with a source located at P (Fig. 1c, right), providing a way to numerically calculate the fields $E^{\text{REV}}(\bar{k}, \bar{\rho}_{SPP})$ by solving the reverse problem.

The probability of a certain outcome is determined by the superposition integral between the forward propagating wave $E^{\text{FWD}}(\bar{k}, \bar{\rho}_{SPP})$ and the time-reversal field $E^{\text{REV}}(\bar{k}', \bar{\rho}_{SPP})$, i.e.,

$$c = \int E^{\text{FWD}}(\bar{k}, \bar{\rho}_{SPP}) E^{\text{REV}}(\bar{k}', \bar{\rho}_{SPP}) d\bar{\rho}_{SPP} . \qquad (4)$$

The partial projection of the total wave function $\Psi$ for the electron-SPP system (Eq. 1) on the specific SPP outcome is therefore

$$|\psi_e(\bar{\rho}_e))\rangle = \left[\int \Psi \cdot E^{\text{REV}}(\bar{k}; \bar{\rho}_{SPP}) d\bar{\rho}_{SPP}\right] . \qquad (5)$$

The qualitative shape of the $E^{\text{REV}}(\bar{k}; \bar{\rho}_{SPP})$ wave is described in Fig. 1b (right). From a quantitative point of view, $E^{\text{REV}}(\bar{k}; \bar{\rho}_{SPP})$ is defined piecewise in the two domains (left and right sides of the double slit feature). At the slit position, the form of $E^{REV}$ is described by a configuration where the spherical wave, reverse-propagating from the bucket, reduces to two plane waves with wave vectors $\overline{k_{1,2}}$ =|K|$(\widehat{\Delta\rho}_{1,2})$, where $\widehat{\Delta\rho}_{1,2}$ is a unitary vector pointing from the jointly measured point to each slit.

At the slit position, $E^{\text{REV}} \approx \delta(\bar{k} - \overline{k_1}) Rect_b\left(y - \frac{d}{2}\right) + \exp(i\varphi)\delta(\bar{k} - \overline{k_2}) Rect_b\left(y + \frac{d}{2}\right)$, where the phase $\varphi$ depends on the post-selection position with respect to the slits. In the slits, superposition between $E^{\text{REV}}$ and $E^{\text{FWD}}$ is possible only if $\bar{k}$ is assumed to be one the specific values $\overline{k_{1,2}}$. After the SPP state is measured using the bucket, the electron wave becomes a coherent superposition of two pure states

$$|\psi_e(\bar{\rho}_e))\rangle = a_1 c_1 |\psi_e(-\overline{k_1}; \overline{\rho_e})\rangle + a_2 c_2 \exp(i\varphi)|\psi_e(-\overline{k_2}; \overline{\rho_e})\rangle , \qquad (6)$$

where $|\psi_e\rangle$ can be adjusted to account for the SPP diffractive propagation from the slits to the electron-SPP interaction area. When the correlated electron wave propagates in the microscope column following the interaction, it forms a coherent diffraction pattern corresponding to the transfer function T. The joint measurement filters out a specific set of $\bar{k}$ from those present in the source state S and a phase that produces a pure state for the electrons. Shifting the detection point would result in a different shifted diffraction pattern. When a series of detection points is summed incoherently or equivalently, since there is no correlation between them, a standard incoherent electron wave can be constructed, fitting well to a ghost imaging scheme. As in the case of photons, a generalized optical path can be considered starting from the bucket detector, through the slits, towards the interaction region from which the rays follow as electron waves.

Ghost imaging would equally work for any one-dimensional object with transmission function $T(y)$ (in the *x-y* plane) instead of the two slits. For example, it would work for a phase object that imparts an effective phase shift on the electron waves.

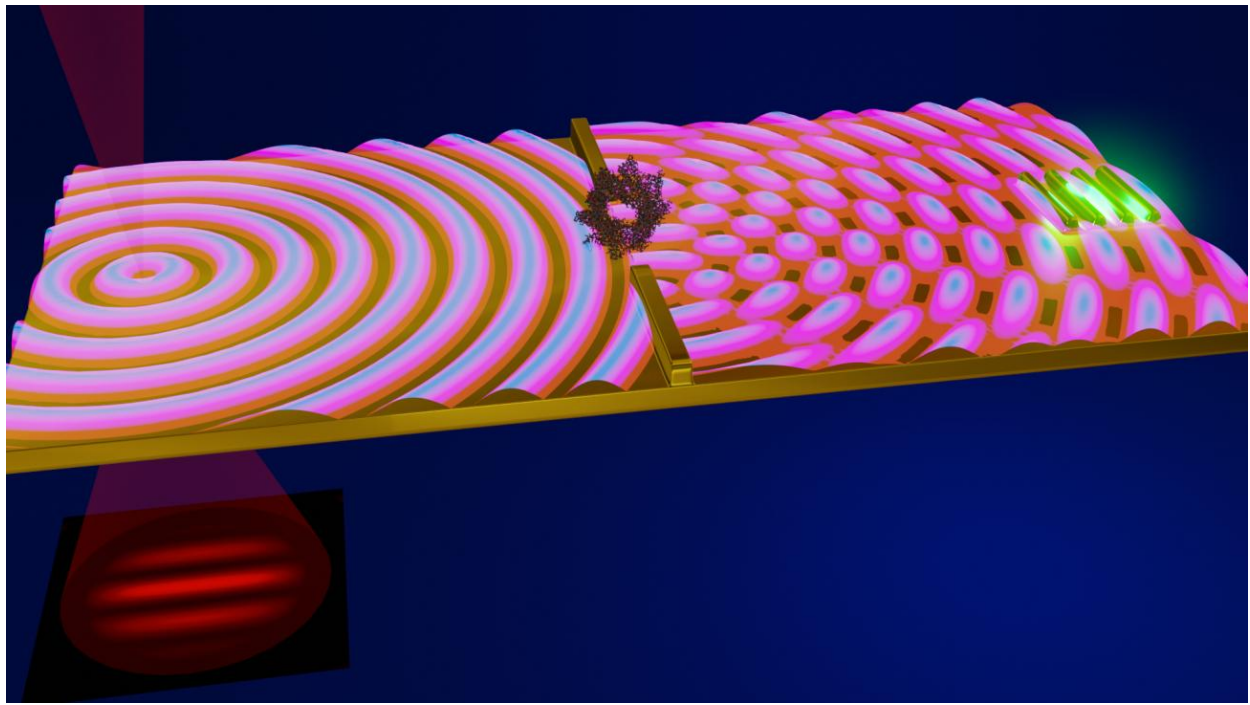

*Figure 2 Schematic diagram of the application of ghost imaging to the imaging of electron-dose-sensitive materials such as biological molecules. The molecules interact with the SPP and the image is reconstructed on the electron beam by performing a joint measurement with the bucket detector on the SPP side.*

Figure 2 shows an application to a relevant case, where the object to be observed is a biological molecule. In such a case, the SPP could excite a degree of freedom of the molecule, resulting in an amplitude effect. It can also result in a local change in the refractive index induced by the molecule as a phase effect. The technique is similar to surface plasmon resonance (SPR), but with very high spatial resolution, together with a well-defined phase and "imaging" sensitivity[40]. The imaging of molecules without damaging them by direct electron beam irradiation, indirectly through an SPP-induced interaction[41], is a very relevant research theme. It would provide an innovative approach for imaging electron-dose-sensitive objects with a spatial resolution that is dictated by the SPP, but also by the distance *D* between the electron beam interaction area and the object (see sup material). It is reasonable to infer that, if D is small, the resolution can be pushed down to D, which can be far below the wavelength of the SPP.

If the SPP scattering from a molecule is only elastic, then the probing is damage-free. Considering the excitation of the molecule through the SPP, the experiment is similar, in terms of dose, to vibrational electron energy loss spectroscopy (EELS) experiments in an "aloof" configuration (i.e., with the electron trajectory not directly on the sample), which reduces damage by orders of magnitude[42]. However, due to the synchronous nature of the measurement, it is natural to expect an even larger information/dose advantage for ghost imaging with respect to normal EELS.

The 1D "ghost imaging" scheme that we propose can also be considered as the reverse of a PINEM mechanism. In PINEM, light excites an SPP mode simultaneously with the electron arrival, such that the interaction between the electron and the local fields driven by the optical excitation modifies or shapes the electron wave function (see Fig. 3). This scheme was used to generate electron vortex beams with structured SPPs[22]. By reversing the coincidence between the electron and the cathodoluminescence light emitted from the electron-driven SPP diffracting from the grating, we find exactly a photon excitation characteristic for the PINEM mechanism to occur (Fig. 3). Reverse PINEM presents several advantages experimentally, since it is not necessary to use a pulsed electron beam, but only to perform a joint measurement to ensure that the electron and photon are originating from the same event. The technical challenge is different than in UTEM – where sub-*ps* control and synchronization are needed. Here, the difficulty lies in the ability to collect light efficiently, to use single photon detectors and to synchronize them with fast cameras to perform joint measurements between electrons and photons, repeating the process for all other spurious events. The recording of joint measurements with a camera, i.e., being able to collect coincident events spatially and temporally, is an existing technology. The rate of electrons emitted per second can be linked to the electron current by the relation, $n = i_e/e$. For a monochromated electron beam, almost all of the inelastic events that are associated with SPP creation correspond to a fraction $P_{SPP}$ of the full electron beam falling on the selected electron energy window. $P_{SPP}$ could also be modified to account for the electron detection efficiency. On average, an electron reaches the EFTEM image every $\tau_{spp} = e/(i_e P_{SPP})$.

Given the use of post selection on a grating, "suitable" photons for coincidence are emitted with probability $P_{PS} = |c|^2|a|^2$, where *a* is the coefficient of k decomposition for SPP generation in Eq. 1 and *c* is the superposition integral for the direct and reverse field in Eq. 4. The factor *a* can be modified to evaluate the coupling strength to the impinging electrons. An approximate evaluation of $|c|^2|a|^2$ can be obtained by considering the fraction of the full 2π angle for which the two slits are observed from the source and from the bucket detector (i.e., the grating). For example, if one assumes that the slits are 1-2 μm in size and located 5 μm from the injection area and 10 μm from the bucket, then $P_B \approx 10^{-3}$ of the coincidence SPP revealed on the bucket.

An effective coincidence experiment should therefore rely on a coincidence window of $\tau_{spp}$ and a dead time before a new acquisition of $\tau_{ps} = e/(i_e P_{SPP} P_{PS})$ (where $\tau_{ps}$ is the average time between two events in a joint measurement). For a current of 10 pA and a $P_{SPP} = 0.1\%$, the required coincidence window is ~10 ns, while the dead time can be as slow as 10 μs. Faster response times, down to few ps, would be advantageous for reducing the dark current in the detectors. The primary limiting factor can be the acquisition time of the camera, which must be on the order of a few ns, which is within the limit of camera technology[43] Coincident experiments in CW have also been demonstrated using delay line detectors[44]. In general, schemes can be based on the acquisition of all EFTEM electrons, with a time stamp placed on each of them, rejecting non-coincident events afterwards.

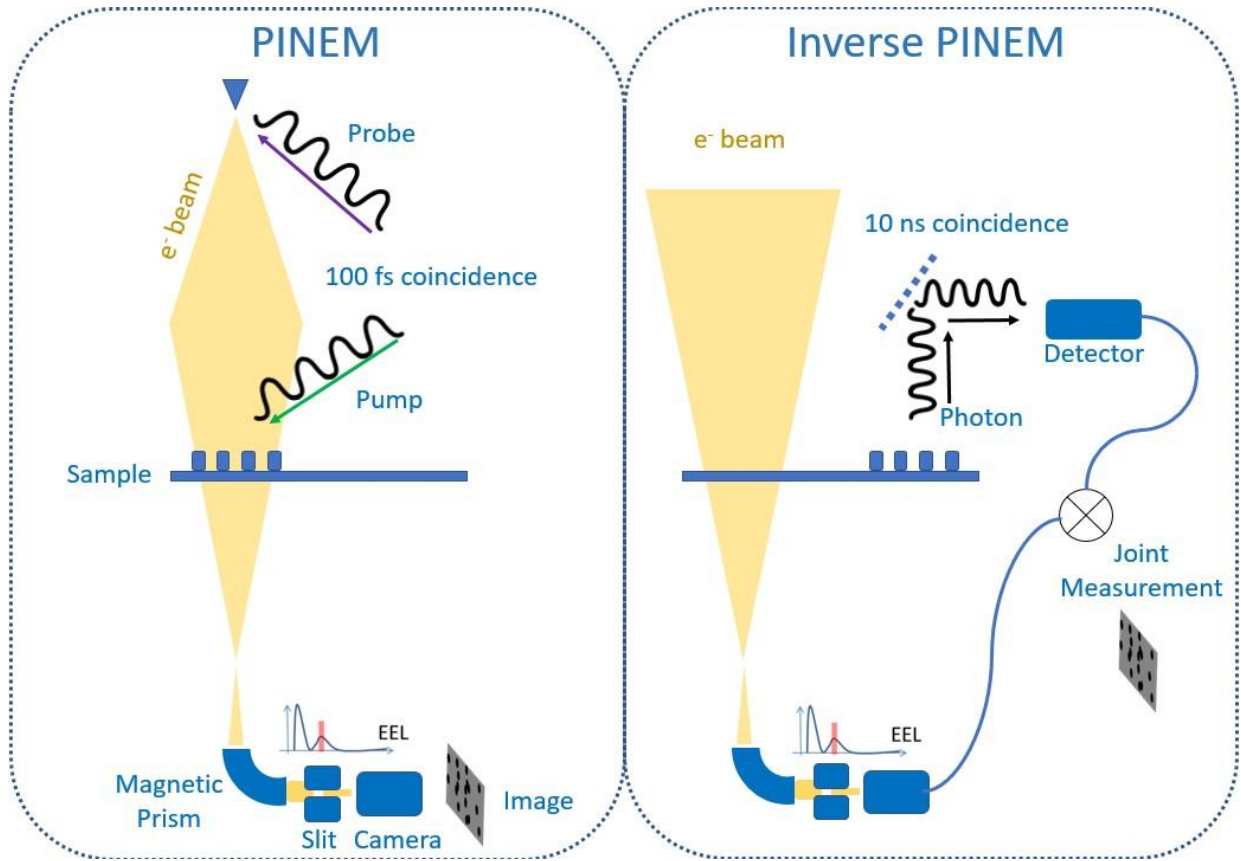

*Figure 3. Comparison between PINEM based on UTEM and inverse PINEM based on one-dimensional ghost imaging.*

In cases for which the detection is slow, the experiment can be improved by shaping the electron wave before interaction to increase the factor $|a|^2$. A variant of the experiment can be performed by placing the slits in the path of the electron beam instead of that of the SPP. The joint measurement can then be performed very similarly to that described above and described according to an inverse-PINEM perspective, with rays starting from the bucket detector and ending in the electron path. While it is well known that, for well-separated slits (e.g., more than 100 nm), the coherence of the inelastically scattered electron is negligible and no interference is expected, a joint measurement permits recovery of the interference figure in the appropriate subset of electrons.

Finally, we take the advantage of the fact that this technique is a reverse version of PINEM to describe its use for dynamic electron beam shaping[25,45]. Instead of changing the electron wavefunction using a light-driven excitation, the electron beam component is singled out by appropriate post selection of the final state. Since more orthogonal selection channels are possible, it is also possible to imagine performing "conditional" beam shaping, whereby the shape of the beam is not decided until the photon is measured. There are many ways to post-select the electron state based on a selected photon state coincidence. In principle, any PINEM-based beam shaping recipe can be applied in reverse. For instance, one can consider structuring the electron beam by replacing the two slits with more complicated patterns.

Since the process is a reverse version of PINEM, the coupling is the same as for interaction with a single SPP:

$$T_e = J_1(\beta) exp\left(i\, arg(-\beta)\right) \approx \beta^* \qquad (7)$$

$$\beta = \frac{e}{\hbar \omega_0} \int_{-\infty}^{+\infty} E_z(R, z) exp(i\frac{\omega_0 z}{v}) dz , \qquad (8)$$

where $J_1(\beta)$ is a first order Bessel function of the first kind and $arg(.)$ is the argument. Equations 7 and 8 are specialized for the case of small coupling, where a single SPP energy-loss band in the final state of the electron and photon is selected. For such a case, the electron transmission function is the complex conjugate of $\beta$ (here, labeled $\beta^*$). For a "well-behaved" z dependence of the field, the equations restate the considerations of momentum conservation that led to Eq. 1. The detection should be structured in such a way as to filter only the appropriate $E_z$ field with the phase.

As an example of electron beam shaping, we choose electron vortex beam generation and make use of the PINEM recipe from Vanacore *et al.*[25]. By making use of a circular hole that acts as discontinuity in a metal slab, circular polarization of the impinging light beam is transformed into orbital angular momentum of the electron beam, *i.e.*, into a vortex shaped $\beta$.

In our simulation, we substituted a single hole with a series of concentric metallic rings separated by $\lambda_{SPP}$. Figure 4 shows that this geometry has the same effect as a single hole, although it allows for a greater coupling and better energy selectivity.

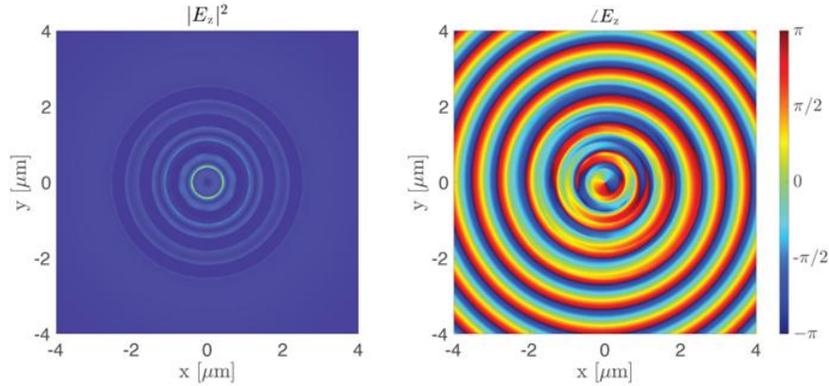

$|Photon \downarrow, Electron \uparrow \rangle + |Photon \uparrow, Electron \downarrow \rangle$

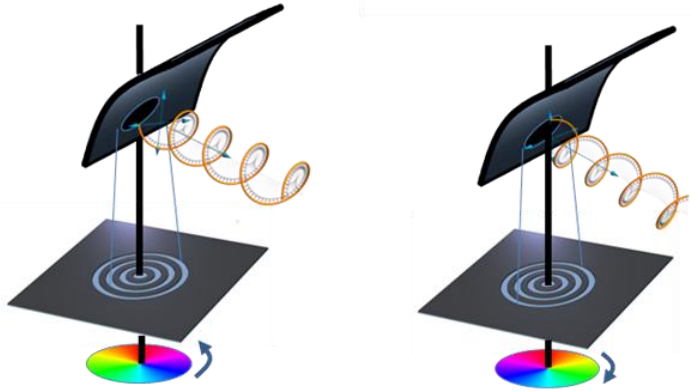

*Figure 4 Example of conditional beam shaping. The structure is a metallic slab patterned with concentric rings. The upper figures shows the phase (right) and the amplitude (left) of the reverse solution. The lower figures show graphical representations of the states Ψ of the systems. The electron beam is in an undetermined state between two opposite vortices. The electron state is defined only after measurement of the photon.*

In the reverse situation, when the electron beam encounters a circular hole, an important fraction of the inelastically scattered part of the electron beam goes into the SPP channel that has a vortex phase modulation (which is a solution of the Maxwell equations for this structure) and becomes correlated with the emitted light. In comparison to the first "1D ghost imaging" experiment discussed above, the grating is replaced by a circular resonator of concentric rings, where the discontinuity allows direct coupling with the unbound radiation. Within this limit, $\ell = +1$ is connected to specific circular polarization whereas $\ell = -1$ is associated with the other. If the slab is positioned in the illumination/condenser region of the microscope, then experiments can be performed with both vortex helicities simultaneously, with a coincidence stamp added to each electron. A further development of this experiment can be considered, where the photon state is not collapsed, but is re-interfered with the beam in different parts of the microscope. This approach is not discussed here.

In conclusion, we have explored a new scheme of imaging based on joint measurement between collective modes produced by electron-specimen interactions and corresponding inelastically-scattered electrons. Such a joint measurement promises to provide a new way to reconsider coherence in inelastic scattering, permitting an increase in the amount of information extracted from each electron incident on the sample. At the same time, coincidence (*e.g.*, between EELS and cathodoluminescence) on the nanosecond time scale with a ghost imaging scheme can replace many photocathode-based UTEM experiments, with advantages in electron beam coherence. The concept of ghost-imaging can be used to examine electron-beam-sensitive materials without the electron beam directly impinging on them. The approach can also be used for beam shaping, with each post selection of a specific photon state corresponding to a different electron beam shape. It promises to open the way to the development of ghost imaging and electron-free interaction imaging techniques in electron microscopy, as well as to explore characteristic quantum phenomena and to verify fundamental quantum inequalities and effects in an electron microscope[46,47].

## Acknowledgments

This project has received funding from the European Union's Horizon 2020 research and innovation programme under grant agreement No 766970 (QSORT), No 964591 (SMARTelectrons), No 101035013 (MINEON) and No. 823717 (ESTEEM3). T. G. and E. K. acknowledge support from the Ontario Early Researcher Award (ERA) and Canada Research Chairs (CRC) Program.

**Supplementary material**

**Ghost imaging resolution**

In standard microscopy based on Abbe's theory, in the far field the resolution is $\Delta r > \lambda$ , where $\lambda$ is the wavelength of the probe.

In ghost imaging, there are two probes (the electron and the SPP) and the resolution is apparently not well defined.

We can define a flat microscope system, in which the main propagation direction is along *x* and the object, just as for the two slits, develops in the *y* direction. We are interested in the resolution in the *y* direction. For example, one may wonder if it is possible to see the 2 slits separated when they are spaced by less than $\lambda_{SPP}$.

As soon as there is SPP propagation at a sufficient distance along *x* from the object, lateral frequencies greater than $1/\lambda$ are not propagated.

This is not entirely correct in the near field. If SPPs are considered to be confined in 2D and monochromatic, then to a large extent

$$k_x^2 + k_y^2 = 1/\lambda_{SPP}^2 \; . \tag{s1}$$

Far-field SPPs are for $|k_{x,y}| < 1/\lambda_{SPP}$ , so the wave is $\exp(i\, k_x\, x + i\, k_y\, y)$. However, there is another class of solutions, which are given for example by $\exp(-\, k_x\, x + i\, k_y\, y)$ , where $|k_y| > 1/\lambda_{SPP}$.

If these evanescent waves can be used, then the resolution can be dictated by the distance D between the electron interaction area and the object.

We consider a "near field" like geometry for the ghost imaging setup. This scheme is shown in the figure below.

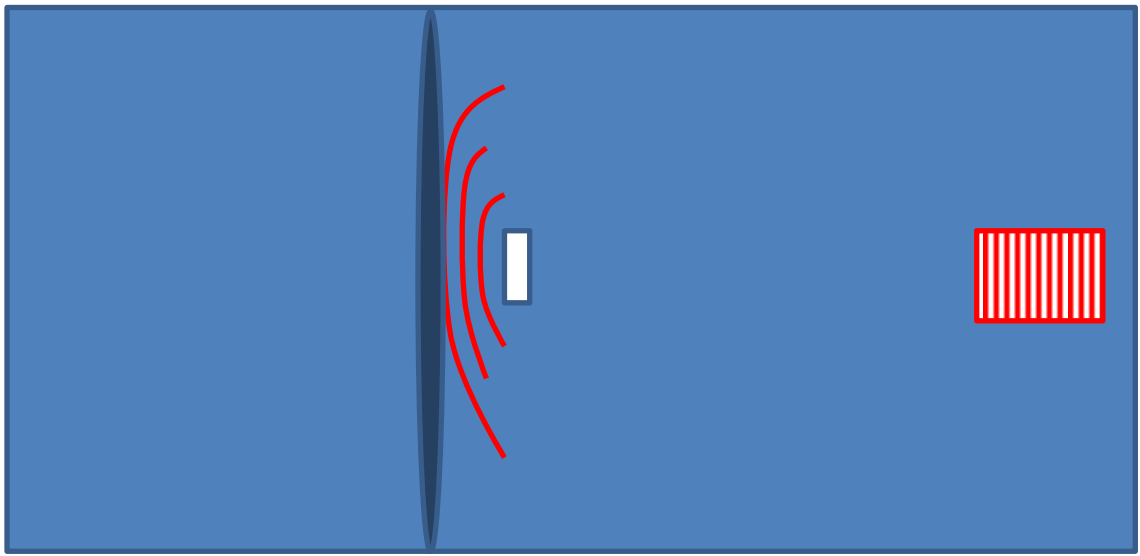

In this scheme, the electron beam is strongly elongated in the *Y* direction and narrow along *X*. The distance D $<\lambda_{SPP}$ can be reduced to 50 nm without inducing knock-on damage on the object.

We now consider the generation area. It produces an electric field as in Eq. 2, convolved with the probe elongated shape.

In 1d Fourier space, the generated field is

$$\widetilde{E_0}(k_y) = \int E_0(y) \exp(ik_y\, y)\, dy \ .$$

where evanescent modes are generated as well.

The object is also characterized by the 1D Fourier development

$$\tilde{T}(k_y) = \int T(y) \exp(ik_y\, y)\, dy \ .$$

The propagation between the two can be modeled by a factor $P_D(k_y)$ that adds a phase effect for low frequencies and an exponential cut roughly as $\exp(-Dk_y)$ to higher frequencies.

After the object, $\widetilde{E_f}(k_y) = \widetilde{E_0}(k_y) P_D(k_y) * \tilde{T}(k_y)$, where * indicates convolution.

The effect of the object's high frequency $k_y$ is to move some of the high frequencies of the injected SPP field to low $k_y$ that are normally propagating to the bucket detector. A high spatial frequency electron will then be observed in the coincidence experiment. The high frequency of the SPP is of course still a very low frequency when compared with the electron wavelength.

If the sample does not have any feature at high frequency, then no high frequency part of the injected SPP will reach the bucket and the electron image will contain no high spatial frequency features.

The ultimate resolution for ghost imaging is therefore expected to be dictated by the distance *D*.